\documentclass[twocolumn,showpacs,preprintnumbers,amsmath,amssymb,prx,floatfix]{revtex4}
\usepackage[pdftex]{graphicx}
\usepackage{dcolumn}
\usepackage{bm}
\usepackage{subfigure}
\usepackage{color}
\usepackage{ulem} 

\begin{document}




\title{%
Lattice dynamics reveals a local symmetry breaking in the emergent dipole phase of PbTe}

\author{Kirsten M. {\O}. Jensen$^{1}$, Emil S. Bo\v zin$^{2}$, Christos D. Malliakas$^{3}$, Matthew B. Stone$^{4}$, Mark D. Lumsden$^{4}$, Mercouri G. Kanatzidis$^{3,5}$, Stephen M. Shapiro$^{2}$, and Simon J.~L. Billinge$^{2,6,*}$}

\affiliation{$^1$Center for Materials Crystallography, Department of Chemistry and iNano, Aarhus University, Denmark}
\affiliation{$^2$Condensed Matter Physics and Materials Science Department, Brookhaven National Laboratory, Upton,~NY~~11973, USA}
\affiliation{$^3$Department of Chemistry, Northwestern University, Evanston,~IL~~60208, USA}
 \affiliation{$^4$Neutron Scattering Science Division, Oak Ridge National Laboratory, Oak Ridge,~TN~~37831, USA}
\affiliation{$^5$Materials Science Division, Argonne National Laboratory, Argonne,~IL~~60439, USA}
\affiliation{$^6$Department of Applied Physics and Applied Mathematics, Columbia University, New York,~NY~~10027, USA}

\date{\today}

\date{}





\begin{abstract}
Local symmetry breaking in complex materials is emerging as an important contributor to materials properties but is inherently difficult to study.  Here we follow up an earlier structural observation of such a local symmetry broken phase in the technologically important compound PbTe with a study of the lattice dynamics using inelastic neutron scattering (INS).  We show that the lattice dynamics are responsive to the local symmetry broken phase, giving key insights in the behavior of PbTe, but also revealing INS as a powerful tool for studying local structure.  The new result is the observation of the unexpected appearance on warming of a new zone center phonon branch in PbTe.  In a harmonic solid the number of phonon branches is strictly determined by the contents and symmetry of the unit cell.  The appearance of the new mode indicates a crossover to a dynamic lower symmetry structure with increasing temperature.  No structural transition is seen crystallographically but the appearance of the new mode in inelastic neutron scattering coincides with the observation of local Pb off-centering dipoles observed in the local structure.
The observation resembles relaxor ferroelectricity
but since there are no inhomogeneous dopants in pure PbTe this anomalous behavior is an intrinsic response of the system. We call such an appearance of dipoles out of a non-dipolar ground-state ``emphanisis" meaning the appearance out of nothing. It cannot be explained within the framework of conventional phase transition theories such as soft-mode theory
and challenges our basic understanding of the physics of materials.

\end{abstract}

\maketitle

\section{Introduction}
The unexpected appearance, on warming, of local Pb off-centering  dipoles was recently reported in PbTe~\cite{bozin;s10}.
No structural transition is seen in the average rock-salt crystal structure but is apparent in the local structure on warming above 100~K: the local symmetry is lowered, losing its centro-symmetry, on warming. We refer to this as ``emphanisis" meaning the appearance of something from nothing since it is fundamentally different from a normal ferroelectric transition where dipoles exist at low temperature but become disordered and fluctuating at high temperature.  Here the dipoles \emph{appear} on warming from a ground-state with no dipoles.  In the original study~\cite{bozin;s10} the dipoles were deemed to be fluctuating but this could not be determined from the experiment itself.  A recent inelastic neutron scattering (INS) study~\cite{delai;nm11} noted the anharmonicity of certain phonons down to low temperature in PbTe.  Here we present a detailed temperature-dependent INS study of the phonons in the temperature range where the local dipoles appear~\cite{bozin;s10}.  Below room temperature our results are in good agreement with earlier work on PbTe~\cite{cochr;prsl66,alper;pl72} and we also see the significantly anharmonic signal of the zone center transverse optical~(TO) mode reported by Delaire~{\it et al.}~\cite{delai;nm11}.  However, the main result of this work is the characterization of the anharmonic features at $\sim 6$~meV  as a new mode with a highly anomalous temperature dependence, growing rapidly in spectral weight with increasing temperature above 100~K, at the expense of the normal TO mode that is the incipient ferroelectric mode~\cite{alper;pl72}.  The new mode and the original TO mode coexist over the entire temperature range measured to 600~K, resulting in an additional phonon branch in the Brillouin zone indicative of a broken symmetry, though none is seen in the average structure. The new mode is broad with a short lifetime, but dispersive. It also hardens with increasing temperature.  Since it appears over the same temperature range where local dipoles appear in the structure~\cite{bozin;s10} we associate its appearance with the appearance of these objects that break the local, though not the average, centro-symmetric symmetry.  Such behavior, observed in a pure binary alloy system, has not been described before and challenges our current understanding of the physics of materials.

\section{Results}
\begin{figure*}[h]
\center
\includegraphics[width=0.75\textwidth,angle=0]{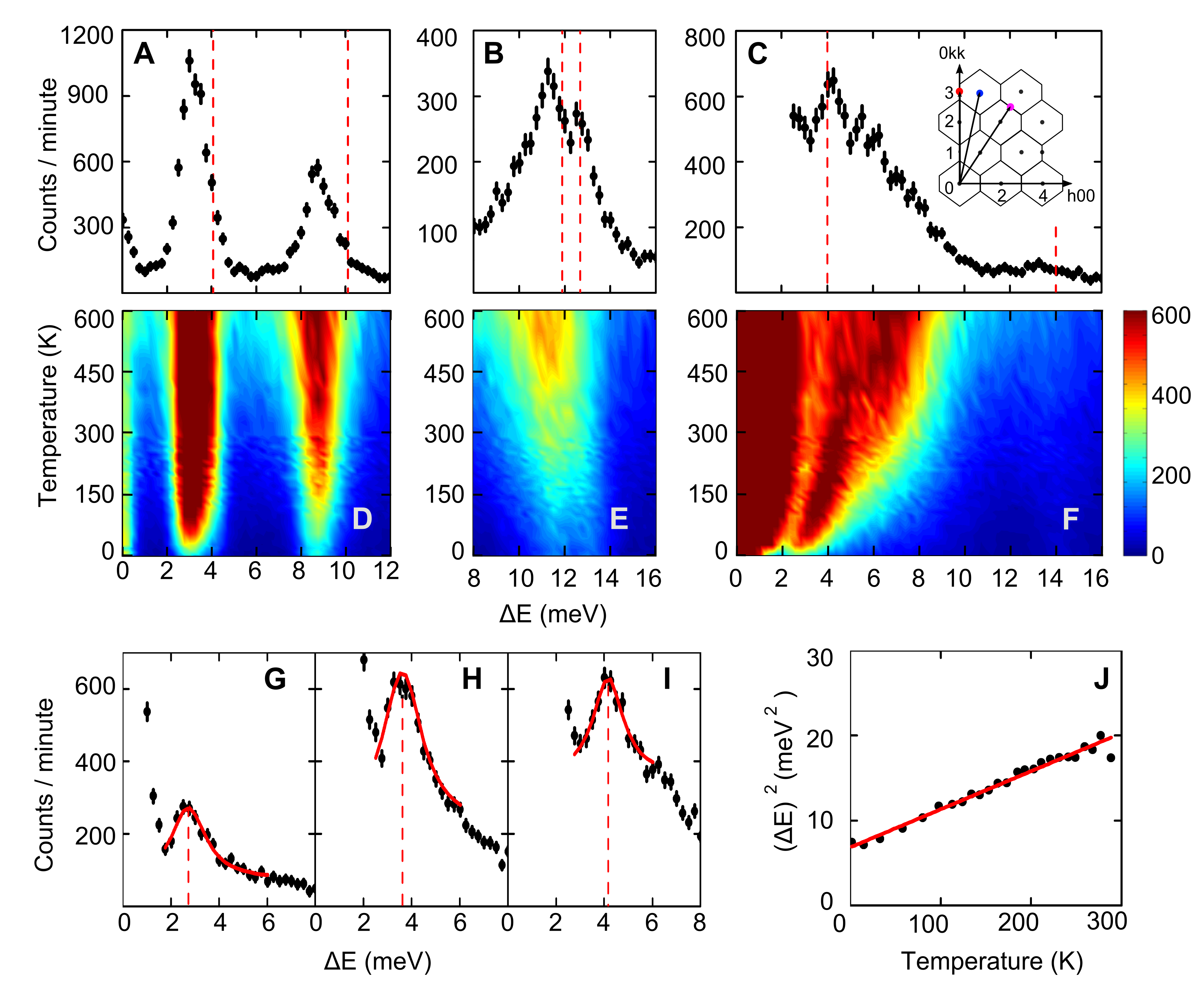}
\caption{
\normalfont (Color) INS spectra of a PbTe single crystal at 288~K at different $\vec{Q}$ points in reciprocal space: { (a)}
(0 3 3) zone edge, { (b)} (2.5 2.5 2.5) zone edge, and { (c)} (1 3 3) zone center. The red dashed lines indicate
the phonon energies measured by Cochran~{\it et. al}~\cite{cochr;prsl66} for comparison.  The inset in { (c)} is a schematic of the plane of reciprocal space that the triple axis spectrometer was set to work in. The grey dots are the reciprocal lattice points and the lines are Brillouin zone boundaries. The red, pink and blue dots are the $\vec{Q}$ points for the data shown in
{ (a)} ($K$-point), { (b)} ($L$ point) and { (c)} ($\Gamma$-point), respectively.  { (d)}-{ (f)}
show false color plots of the intensity (color axis) vs. energy transfer, $\Delta E = \hbar\omega^2$, and temperature of the same modes shown in the panels above. { (g)}-{ (i)} show representative fits of Lorentzian peaks (in red) to the TO mode at the [133] zone center spectra: { (g)} 2~K, { (h)} 134~K and { (i)} 231~K. The vertical dashed lines are the extracted mode energies, $\omega$. (j) $\omega^2$ vs. $T$ for all the temperatures fit in the temperature range below 300~K with a straight line fit to the data shown in red. }
\label{fig;lowT}
\end{figure*}

We first consider the lattice dynamics at room temperature and below, and compare our measurements to earlier results in the literature~\cite{cochr;prsl66,alper;pl72,delai;nm11} to establish the quality of our sample and data.  Fig.~\ref{fig;lowT}(a)-(c) shows representative room temperature INS constant $\vec{Q}$ scans, at three points in the Brillouin zone, collected on the HB3 triple axis spectrometer at Oak Ridge National Laboratory (ORNL). Details of the data collection are described in the Materials and Methods section below.
As illustrated in the inset to Fig.~\ref{fig;lowT}, the points in reciprocal space are (a) (033), which is a zone boundary $K$-point~\cite{dress;b;gtattpocm08} in a position such that the instrument is sensitive to transverse polarized modes~\cite{shira;b;nswatas02} (b) (2.5 2.5 2.5), which is a zone boundary $L$-point sensitive to longitudinal modes, and (c) (133), which is a Brillouin zone center ($\Gamma$-point) where both longitudinal and transverse modes can be measured.  Certain phonon branches are expected based on the earlier work of Cochran~{\it et. al}~\cite{cochr;prsl66}, the energy transfers of which are indicated by red dashed lines in Fig.~\ref{fig;lowT}.  There is good semi-quantitative agreement with the earlier results. Fig.~\ref{fig;lowT}(d)-(f) shows a false-color plot of the scattered intensity as a function of temperature and energy transfer, $\Delta E = \hbar\omega$, for the same three $\vec{Q}$-points.  The $K$-point modes in Fig.~\ref{fig;lowT}(d) do not change energy at all vs. temperature, neither do they broaden significantly.  This is expected in a harmonic system, but even in a quasi-harmonic approximation where lattice expansion is taken into account it is common to see softening with increasing temperature~\cite{fultz;pms10}.  The LA/LO doublet at the $L$-point shown in Fig.~\ref{fig;lowT}(e) does appear to broaden with increasing temperature though the peak contains two components that are both softening with increasing temperature.  Fits to the line-shapes suggest that there is little broadening of the individual lines.  Qualitatively, the softening of these modes may be explained by a quasi-harmonic analysis.

We now consider the zone-center TO mode in Fig.~\ref{fig;lowT}(c) and (f).  This mode is the soft-mode that indicates an incipient ferroelectric phase transition in PbTe~\cite{alper;pl72}.  The mode softens with decreasing temperature (the opposite of the $L$-point modes and opposite to early density functional theory (DFT) calculations~\cite{an;ssc08}) but never reaches zero frequency.  A softening to zero frequency would indicate the soft-mode transition temperature~\cite{bruce;b;spt81} and a structural phase transition.  It is immediately apparent that the mode is anomalously broad as pointed out recently~\cite{delai;nm11}.  The zone edge modes in Fig.~\ref{fig;lowT}(a) and (b) have a FWHM of around 1.5~meV which is close to the calculated energy resolution of the instrument, as described in Materials and Methods below.  However, the high-energy tail on the zone center TO mode is of order~$\sim 5$~meV.  The  temperature dependence of this scattering feature is shown in a false-color plot in Fig.~\ref{fig;lowT}(f).  This feature in the scattering is seen to broaden dramatically with increasing temperature and is highly asymmetric and non Gaussian~\cite{delai;nm11}.

To study the temperature dependence of this mode in greater detail we have fit curves to the spectra at each temperature.  We focus initially on the low-temperature region, $T<300$~K.  We would like to see if our data reproduce the earlier results of Alperin {\it et al.}~\cite{alper;pl72}.  Reasonable fits could be made to the scattering features by fitting a Lorentzian peak on a linear sloping background, and from this we extracted the mode frequency. Three of these fits are shown in Figs.~\ref{fig;lowT}(g)-(i). The temperature dependence is shown in Fig.~\ref{fig;lowT}(j).  For a mean-field, second order phase transition, the soft mode theory predicts that the frequency of the soft mode approaches zero as $(T-T_c)^{0.5}$~\cite{bruce;b;spt81}, so a plot of $\omega^2$ vs. $T$ will be linear and intercept the abscissa at transition temperature, Tc.  It is clear that the zone center mode softens on cooling according to mean-field soft mode theory but still has a positive frequency at $T=0$: PbTe is an incipient ferroelectric but, in the absence of alloying with Ge~\cite{jants;b;dpoivc83}, does not undergo a ferroelectric phase transition on cooling.  There is excellent agreement with the earlier work~\cite{alper;pl72}.  Extrapolating the line to the abscissa we obtain a putative phase transition temperature of -151~K.  This is a little more negative in temperature than Alperin~{\it et al.}~\cite{alper;pl72} (-135~K) but in good general agreement and establishes the purity of our crystal.

\begin{figure}[h]
\center
\includegraphics[width=0.8\columnwidth,angle=0]{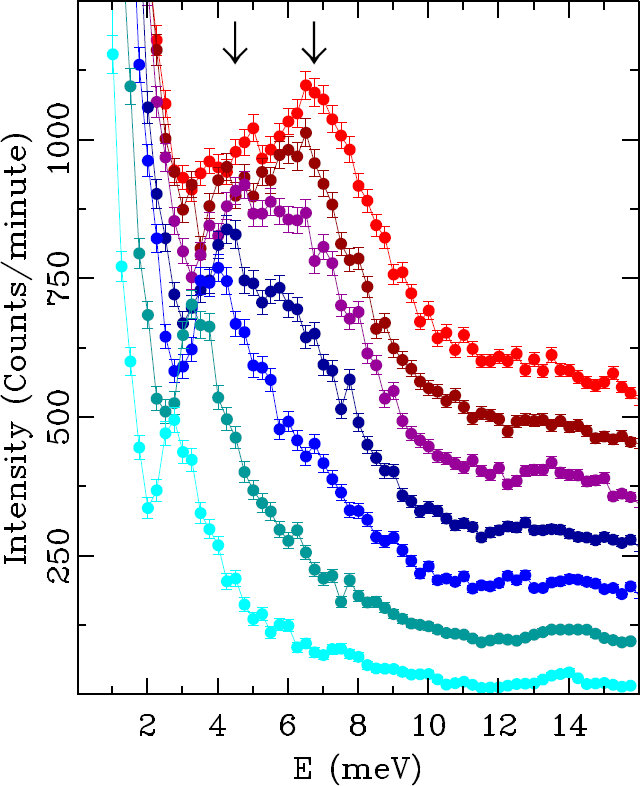}
\caption{
\normalfont (Color) Plots of the INS spectrum at $\vec{Q}=(133)$.  From
the bottom (pale blue) the
temperatures shown are 28~K, 99~K, 203~K, 299~K, 410~K, 486~K, and 601~K (topmost curve). Datasets are offset
with respect to each other by 75 intensity units for clarity. The two arrows are guides to the eye,
indicating the positions of the TO and the new mode at the highest temperature.  Both modes soften on
cooling.}
\label{fig;newMode}
\end{figure}

Having established that data from our sample are in good agreement with earlier work we move to the main result of the current study.  In Fig.~\ref{fig;newMode} we show the temperature dependence of the zone center TO mode feature at and above room temperature.
Over a narrow temperature range beginning above $\sim 100$~K a new peak grows up rapidly but smoothly from the high energy tail of the TO mode, gaining spectral weight at the expense of the TO mode, but coexisting with it. Above room temperature the new peak becomes very strong, dominating the original TO feature.  It indicates the presence of a new mode in the system that appears on increasing temperature, which is highly anomalous.

The neutron scattering dynamic structure factor, $S(\vec{Q},\omega)$, is the normalized inelastic scattering intensity and is related to the imaginary-part of the dynamic susceptibility according to~\cite{shira;b;nswatas02}
\[S(\vec{Q},\omega) = \frac{\chi^{\prime\prime}(\vec{Q},\omega)}{(1-e^{-\frac{\hbar\omega}{k_BT}})}.\]
For a particular phonon mode in the $s$th branch at position $\vec{Q}=\vec{G}-\vec{q}$ in reciprocal space, where $\vec{G}$ is a
reciprocal lattice vector and $\vec{q}$ is a vector that lies within a single Brillouin zone, $\chi^{\prime\prime}(\vec{Q},\omega)$ is proportional to, $|F(\vec{Q})|^2/\omega_{\vec{q}s}$, where
$F(\vec{Q})$ is the phonon dynamic structure factor.  It is proportional to the Bragg scattering structure factor for zone center acoustic modes, but in general depends on the polarization of the vibrational mode and the atomic displacement vectors of the eigenfunction.

\begin{figure*}[h]
\center
\includegraphics[width=0.75\textwidth,angle=0]{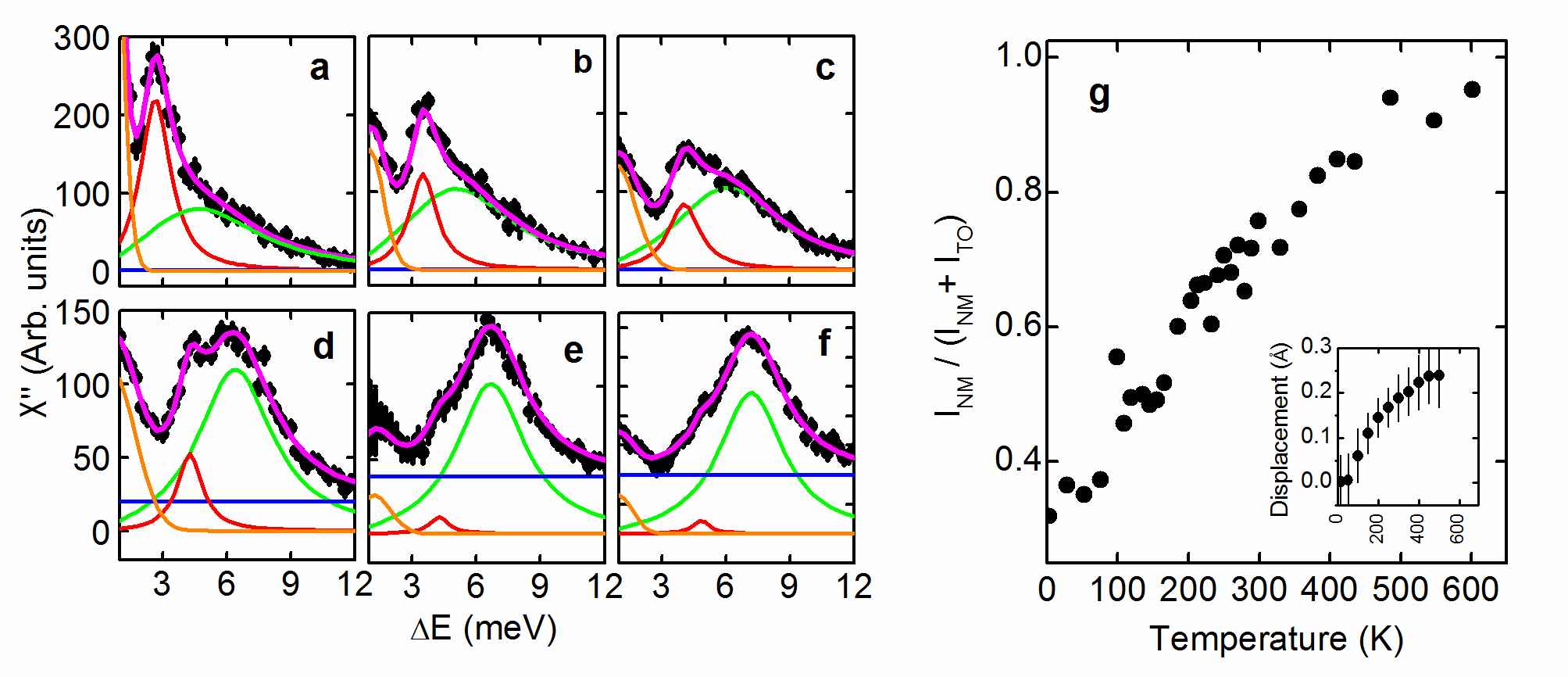}
\caption{
%
\normalfont (Color) { (a)}-{ (f)} : Dynamic susceptibility obtained from the measured INS spectra (symbols) at representative temperatures with fits to the data shown as a pink line. There are 2 damped harmonic oscillator components, one for the TO mode (red) and one for the new mode (green). A Gaussian function was added to fit the intensity from the tail of the Bragg peak (orange). The horizontal blue line is a constant background component. { (a)} 2 K, { (b)} 98 K, { (c)} 203 K, { (d)} 299 K, { (e)} 482 K and { (f)} 601 K. { (g)} shows the temperature dependence of the integrated weight in the new mode from the fits (black symbols). The inset shows the Pb off-centering displacement measured from the PDF measurement on PbTe reported in~\cite{bozin;s10}
}
\label{fig;nmfits}
\end{figure*}

\begin{figure*}[h]
\center
\includegraphics[width=0.75\textwidth,angle=0]{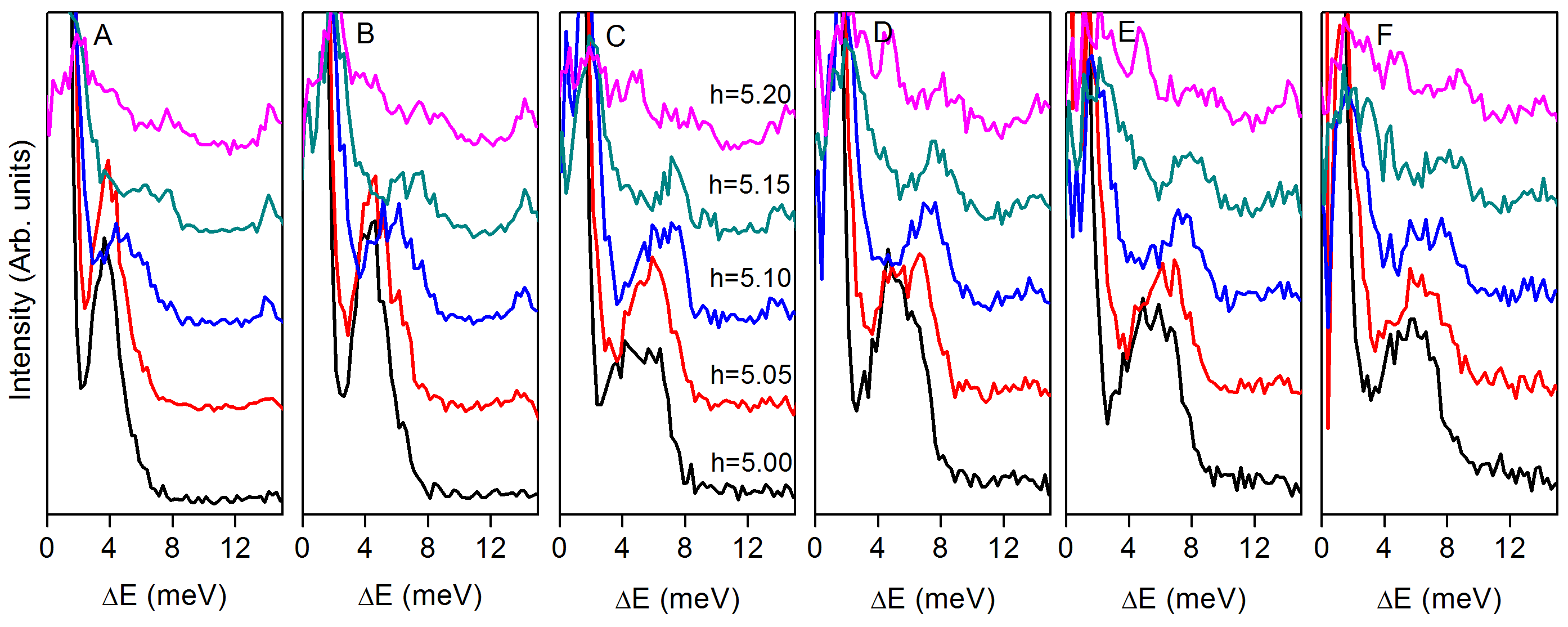}
\caption{
(Color) Neutron scattering intensity vs. energy transfer from single crystal PbTe measured on the ARCS spectrometer. Curves in each panel are cuts along $[h11]$ for 5 (zone center)$ < h < 5.2$ showing the dispersion of the modes.
\normalfont {(a)} 100~K, { (b)} 200~K, {(c)} 300~K, {(d)} 400~K, {(e)} 500~K and {(f)} 600~K.
At high temperature the intensity is almost exclusively coming from the new mode which is seen clearly dispersing
to higher energy with increasing $h$. }
\label{fig;dispersion}
\end{figure*}

To obtain a quantity proportional to $\chi^{\prime\prime}(\vec{Q},\omega)$, we multiply the measured intensity by the Bose-Einstein factor $C= (1-e^{-\frac{\hbar\omega}{k_BT}})$ .  This removes the temperature dependent effects of phonon occupancy and allows us to track the changes in the phonon energy spectrum directly . The thus normalized datasets from $\vec{Q}=[331]$ were then fit with 2 damped harmonic oscillator functions to extract the different phonon dynamic susceptibilities. An additional Gaussian function was introduced to fit the intensity from the tail of the Bragg peak. Details of the fitting are in the Materials and Methods section. Representative examples of the fits to the data at various temperatures are shown in Fig.~\ref{fig;nmfits}(a)-(f).
The growth of the susceptibility of the new mode, at the expense of the TO mode, with increasing temperature is clearly apparent, as evident in Fig.~\ref{fig;nmfits}(g). As well as growing, the new mode sharpens and hardens with increasing temperature. Rapid growth of the new mode spectral weight begins at around 100~K. The behavior is reminiscent of the power law growth of an order parameter in a Ginsberg-Landau theory of a conventional second order phase transition~\cite{bruce;b;spt81} except in that case the order parameter grows upon cooling through the phase transition, not on warming as here.  The temperature dependence of the mode susceptibility is reminiscent of the growth in the amplitude of the off-center Pb displacements observed in the atomic pair distribution function (PDF)~\cite{bozin;s10}, as shown in the inset. We have investigated the dispersion of the new mode in the $[h11]$ direction away from the zone center using the ARCS chopper spectrometer at ORNL.  As is  evident from Fig.~\ref{fig;dispersion},
the mode clearly disperses to higher energy transfer moving away from the zone center, similar to the TO mode, and hardens with increasing temperature.

The observation of additional modes implies a symmetry breaking in the system.  The most trivial possibility is the appearance of a localized mode due to the presence of  point defects in the material. Such modes are extremely weak and non-dispersive but can become apparent, even for modest defect densities, when they interact with dispersing crystal modes~\cite{nickl;prb79}.  This explanation can be ruled out by the temperature dependence.  100~K is too low a temperature for any significant thermally activated defect formation and it is difficult to explain the power-law behavior of the susceptibility in this scenario. Also, the new mode is dispersive arguing against this explanation.  Similar arguments argue against another possibility which is the formation of thermally induced intrinsic local modes~\cite{manle;prb09}.  These are localized phonon modes that form due to non-linear effects in a strongly driven system of oscillators~\cite{campb;pt04} and have been postulated to form due to thermal excitation in soft ionic systems~\cite{manle;prb09}.  An anharmonic coupling between the LA and TO modes has also been postulated to explain the breadth of the TO mode in PbTe~\cite{delai;nm11}, but this does not explain the characteristic temperature dependence, and the clear observation of two distinct coexisting components in the vicinity of the TO mode. Also the LA mode is at zero frequency and far from the TO mode at the zone center where our data were measured.
The most striking correspondence of the new mode is with the temperature dependent appearance of Pb off-centering  structural dipoles in the local structure reported from PDF measurements~\cite{bozin;s10}. This observation was rationalized on thermodynamic grounds as the entropically stabilized appearance of a paraelectric phase above an undistorted non-ferroelectric ground-state.  In this picture, the emergent dipoles break the local symmetry by removing the center of symmetry on the Pb site, even though there is no change in the long-range symmetry.  This will result in short and long Pb-Te bonds, changing mode frequencies associated with the TO vibrations which are the intra-unit cell anti-phase vibrations of Pb and Te, just as observed here.  If each off-center Pb atom is fluctuating independently, this would result in a flat non-dispersing mode.  The observation of dispersion implies that off-centered Pb ion displacements are correlated over some range of space, resulting in polar nanodomains similar to those observed in relaxor ferroelectrics below the Burns temperature~\cite{burns;prb76}. The new mode seen here is highly reminiscent of modes appearing in INS measurements of relaxor ferroelectrics that were attributed to polar nanodomains~\cite{vakhr;prb02}. However, the crucial distinction here is that PbTe is  a pure material.  In relaxors the polar nanodomains are stabilized by nanoscale chemical disorder whereas here they must be intrinsic. A physical explanation may lie in the partially localized optical phonon modes recently postulated byDFT and molecular modeling~\cite{zhang;prl11} that result from partial covalency of the Pb and Te that competes with the electrostatic Madelung potential which prefers the high symmetry rock-salt structure. More work on the temperature dependence and the dispersion of these modes is needed to establish the true origin of this behavior.

\begin{figure*}[h]
\center
\includegraphics[width=0.75\textwidth,angle=0]{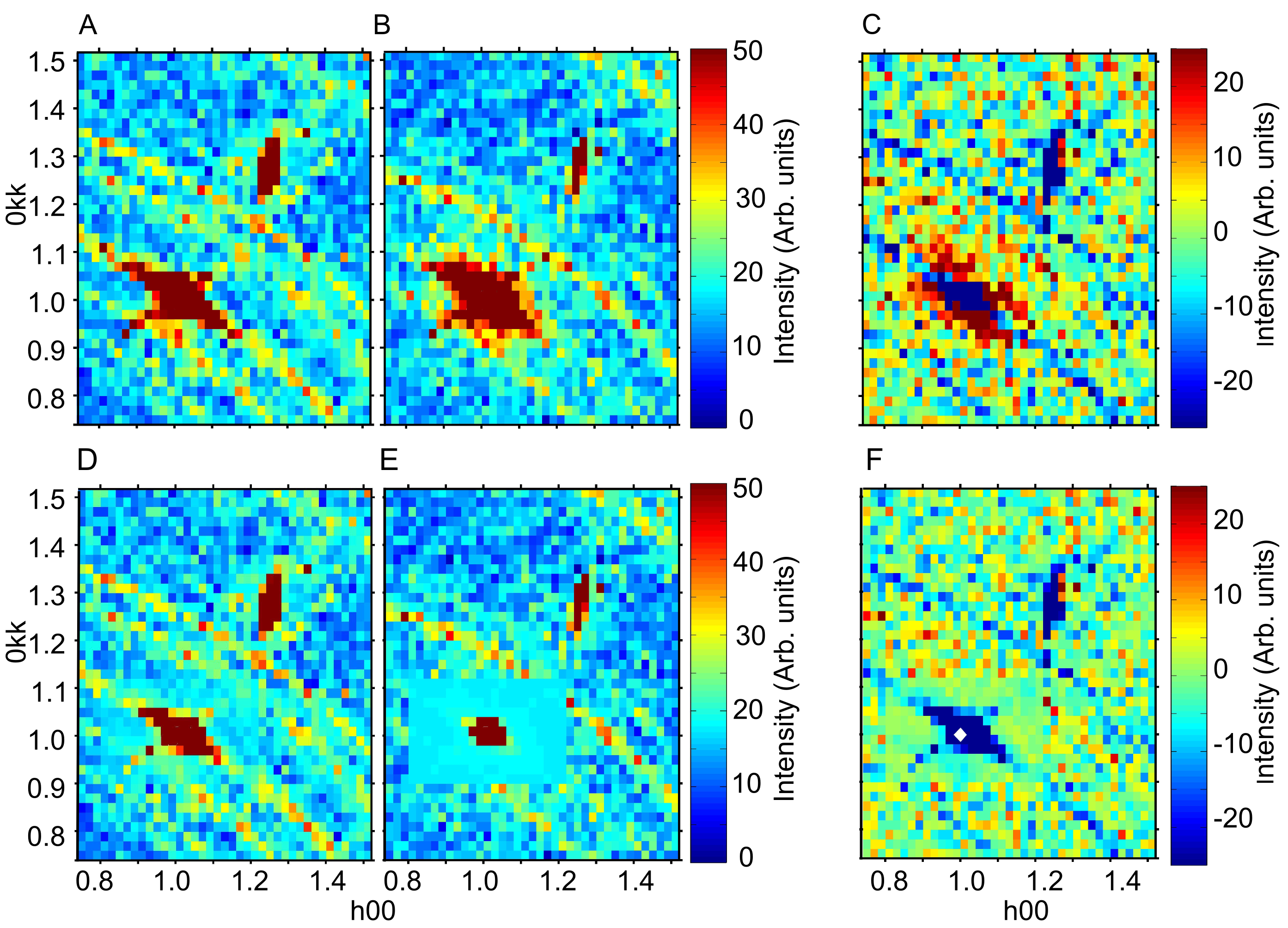}
\caption{
(Color) Elastic diffuse scattering around the [h00],[0k1] plane of reciprocal space
measured using the Triple axis spectrometer:
\normalfont { (a)} 16~K, { (b)} 482~K { (c)} difference between the scans in { (a)} and { (b)}: $I(482)-I(16)$.
{ (d)} and { (e)}, as { (a)} and { (b)} but the intensities have been corrected to account for the change
in phonon occupation of the acoustic modes that are within the energy resolution of the measurement. { (f)} The difference between the scans in { (d)} and { (e)}}
\label{fig;elasticDiffuse}
\end{figure*}

We would like to establish if there is a static component to the short-range ordered nanoscale distortions by searching for diffuse scattering around the zone center in the elastic channel ($\omega = 0$). Indeed, we find there is a small increase in scattering at the base of the (111) Bragg peak at high temperature, which could be elastic diffuse scattering, as evident by comparing Figs.~\ref{fig;elasticDiffuse}(a) and (b).  The Bragg peak is evident as the strong scattering signal in dark red centered at (1.0,1.0,1.0).  Another strong feature at (1.25,1.3,1.3) is a spurious feature of unknown origin.  Powder diffraction lines are also evident coming from the aluminum cryostat windows, and for the line going through the Bragg peak, possibly some crystal mosaicity. The spurious features disappear (or rather appear with negative differential intensities) in the difference plot in Fig.~\ref{fig;elasticDiffuse}(c) which shows the difference between the intensities at high, (b), and low, (a), temerature.  We see some diffuse scattering intensity appearing at high temperature in this elastic measurement, suggesting that the high temperature broken symmetry phase has a static component.  This extra diffuse scattering component shows up as positive intensity around the (111) point in Fig.~\ref{fig;elasticDiffuse}(c)
whereas the Bragg features have a negative difference (dark blue) due to Debye-Waller effects.
However, the most likely origin of the observed increase in elastic diffuse intensity is the finite energy resolution of our measurement resulting in a temperature dependent signal coming from the increase in population of the acoustic phonons that lie within the energy resolution window.  Fig.~\ref{fig;elasticDiffuse}(d) and (e) show scans that have been approximately corrected for phonon effects by dividing the region around the Bragg peak by the Bose factor appropriate for the phonon modes in the resolution window.  This correction accounts for all of the observed diffuse scattering at high temperature in Fig.~\ref{fig;elasticDiffuse}(c) (see Fig.~\ref{fig;elasticDiffuse}(f)) suggesting that there is \emph{no elastic diffuse scattering signal} from the dipoles and therefore they have no static component but are completely dynamic in nature, again consistent with the DFT prediction~\cite{zhang;prl11}

The chemically simple binary PbTe compound continues to turn up surprises that challenge our understanding of condensed matter systems.  We know of no other example of the appearance in a pure material of a new phonon mode with increasing temperature, and the characteristic temperature dependence of its dynamic susceptibility,  argues strongly that this is an intrinsic response of the system.  The inelastic scattering is sensitive to the local symmetry which is broken at high temperature where the new mode is seen to coexist over a wide range of temperature with the original TO mode of the undistorted structure.  A picture emerges of nanoscale regions where the local dipoles appear at high temperature and are correlated and fluctuating.  This lowering of local symmetry with rising temperature, betrayed by the emerging modes seen here and in the PDF~\cite{bozin;s10}, may explain the long known anomalous temperature dependence of the semiconducting energy gap in PbTe~\cite{taube;jap66,zhang;prl11}. Unlike in conventional semiconductors where the energy gap is known to increase with falling temperature due to the reduction of the intensity of thermal vibrations, that of PbTe increases with increasing temperature up to as high as 600~K. It is this anomalous dependence that permits PbTe to exhibit delayed cross gap carrier excitations thereby sustaining a very high thermoelectric power factor at high temperatures. Coupled with the increased anharmonicity and phonon scattering from the same origin resulting in a very low thermal conductivity, this may help to explain the "one two punch" that propels PbTe to the top of its thermoelectric class~\cite{dugha;pb02}.

\section{Materials and Methods}
\subsection{Sample preparation}

Stoichiometric amounts of Pb (Rotometals, at 99.9\% purity) and Te (Plasmaterials, at 99.999\% purity) were flame sealed in an evacuated ($<10^{-4}$~mTorr) fused silica tube. The mixture was heated to 1050~C at a rate of 100~C/h for 4~h and cooled down to room temperature at a rate of 20~C/h.  For the single crystal growth, $\sim 20$~g of PbTe were remelted using the Bridgman method. A 13~mm fused silica tube was loaded with PbTe and lowered at a rate of $\sim 2.6$~mm/h through a single zone vertical furnace that was set at 1050$^\circ$C. A cylindrical shaped crystal of 13~mm in diameter and $\sim 60$~mm in height was used for the inelastic neutron scattering experiment. The quality of the single crystal was evaluated using neutron Laue diffraction at SNS.

\subsection{Triple Axis Inelastic Neutron Scattering Experiments}
The inelastic neutron experiments were performed at the HB3 triple axis instrument at the High Flux Isotopie Reactor (HFIR) reactor located at Oak Ridge National Laboratory.  All the experiments were performed with a fixed final energy, $E_f$= 14.7~meV using Pyrolytic Graphite (PG) as a monochromator and analyzer and a PG filter after the sample to eliminate higher order contamination of the scattered beam.  The neutron beam collimation was $48'$-$20'$-$40'$-$90'$ before the monochromator, sample, analyzer, and detector, respectively.  This yielded an energy resolution of  1.06~meV full width half maximum (FWHM) for zero energy transfer, increasing to 1.62~meV FWHM at $\omega = 12.0$~meV energy transfer.  The sample was mounted to the cold-tip of a closed cycle He-4 based
refrigerator (CCR).  Data scans were made as a function of energy transfer at constant $\vec{Q}$ points in reciprocal space at 35 different temperatures between 2.5~K and 600~K.

\subsection{Triple Axis Inelastic Neutron Scattering Data Curve Fitting}
Here we describe the method used to extract the dynamic structure factors of the modes at the (133) zone center from the TAS data. First the data are normalized to obtain the dynamic susceptibility, $\chi^{\prime\prime}(\vec{Q},\omega)$, as described in the main paper. Based on earlier work we expect two zone-center modes to be present at finite energy transfer (the acoustic modes are hidden under the elastic line at $\omega = 0$): the TO mode at around 4~meV and the LO mode around 14~meV. As the LO mode does not overlap with either the TO or the new mode, we do not take this into account in the fit, which only includes the region from 1-12~meV. The TO mode is then fitted with a function representing a damped harmonic oscillator~\cite{shira;b;nswatas02}, with an additional constant background in the fit. Furthermore, a Gaussian function is added to take care of the tail of the Bragg peak which is present even after the normalization with the Bose factor. We then add an additional damped harmonic oscillator to account for the new mode. The spectra were then fit as a function of temperature sequentially with the starting model for each fit using the parameters obtained from the previous temperature point. Certain constraints were introduced in order to aid convergence, namely, the widths of TO and new mode curve and the background level were allowed to change by only $\pm 20$\%, between subsequent runs.

\subsection{Triple Axis Elastic Diffuse Scattering Measurements}
Elastic scans were also taken on a grid of points in the Brillouin zone At 16~K and 482~K to search for  diffuse scattering from static disorder.  The range $0.8<h<1.5$ was scanned in the $[h00]$ direction and $0.8<k<1.5$ in the $[0kk]$ direction to capture the region around the (111) reciprocal lattice point.  This is a zone center and the PbTe Bragg peak is evident in Fig.~\ref{fig;elasticDiffuse}(a) at the (111) point. There is an additional Bragg feature at around (1.25 1.3 1.3). We are not sure of the origin of this feature and think it is a spurious peak. It does not show any interesting temperature dependence and we neglect it in the analysis.  In addition there are a number of powder Bragg lines evident.  One goes through the (111) point and suggests some sample mosaicity.  The other powder lines probably come from the aluminum sample environment.

In Fig.~\ref{fig;elasticDiffuse}(d) and (e) we have applied a simple correction to account for the increased occupancy of acoustic phonons that lie within the energy resolution of the measurement.  Based on the dispersion curves measured by Cochran~{\it et. al}., we estimate the range of $\vec{Q}$ that is affected by phonons within the energy resolution (of around 1~meV). We assume that the average phonon energy in that region is $\sim 0.5$~meV and so correct the intensities in this range of $\vec{Q}$ by the Bose factor appropriate for a mode of that energy and the temperature in question.  This is a highly simplified correction and is intended to estimate the scale of intensity coming from the increase mode population at high temperature rather than being a highly accurate correction.  We see that this simple correction accounts for essentially 100\% of the diffuse scattering seen at higher temperature in the elastic channel.

\subsection{Chopper Spectrometer Inelastic Neutron Scattering Experiments}
Inelastic neutron scattering measurements were also performed using the ARCS direct
geometry time-of-flight chopper spectrometer at the Spallation Neutron Source (SNS)
at the Oak Ridge National Laboratory.   Single crystal measurements were performed
with an incident energy of
$E_i=30$~meV with the sample mounted approximately in the $(hk0)$ scattering plane
of the instrument.
The sample was mounted to the cold-tip of a closed cycle He-4 based
refrigerator (CCR).
Data were acquired by rotating the sample about the vertical axis by 40 degrees, and
collecting data in 1.875 degree increments.  Empty sample can measurements were
performed for all data acquired at ARCS and subtracted from the data shown in
the manuscript.

The resulting large 5-dimensional data-sets show the scattering intensity throughout
much of $(\vec{Q},\omega)$ space.  The data are projected into a
4D space and 3D slices and 2D cuts are made to extract features of interest using
the DCS-mslice program within the DAVE~\cite{azuah;jrnist09} software package.

\acknowledgments{
We would like to thank Simon Johnsen for help with growing the crystal and Nicola
    Spaldin and Petros Souvatzis for helpful discussions.
    Work in the Billinge group was supported by the Office of Science, U.S. Department of Energy (OS-DOE), under
    contract no. DE-AC02-98CH10886.
    Work in the Kanatzidis group was supported as
part of the Revolutionary Materials for Solid State Energy Conversion,
an Energy Frontier Research Center funded by the U.S. Department of
Energy, Office of Science, Office of Basic Energy Sciences under Award
Number DE-SC0001054.
    The neutron scattering measurements were carried out at the HFIR and SNS at
    Oak Ridge National Laboratory was sponsored by the Scientific User Facilities
    Division, Office of Basic Energy Sciences, U. S. Department of Energy.}




\end{document}